# Doppler-free Fourier transform spectroscopy

Samuel A. Meek,[1] Arthur Hipke,[1,2] Guy Guelachvili,[3] Theodor W. Hänsch[1,2] and Nathalie Picqué[1,2,3*]

1. Max-Planck-Institut für Quantenoptik, Hans-Kopfermann-Straße 1, 85748 Garching, Germany
2. Ludwig-Maximilians-Universität München, Fakultät für Physik, Schellingstr. 4/III, 80799 München, Germany
3. Institut des Sciences Moléculaires d'Orsay (ISMO), CNRS, Univ. Paris-Sud, Université Paris-Saclay, F-91405 Orsay, France
*Corresponding author: nathalie.picque@mpq.mpg.de

**The feasibility of sub-Doppler broadband multi-heterodyne spectroscopy with two laser frequency combs is demonstrated with two-photon excitation spectra of the 5*S*-5*D* transitions of rubidium vapor.**

Fourier transform spectroscopy (FTS) has been for nearly fifty years the leading technique in analytical chemistry and molecular spectroscopy, as well as an irreplaceable tool for remote sensing, industrial process control etc. Fourier transform spectrometers record on a single photo-detector, in almost any spectral region, multiplex high-resolution spectra over a broad spectral span. Interferometers with a resolution better than 30 MHz are available, even commercially; however high resolution FTS in the gas phase has been so far limited to the measurements of Doppler-broadened lines, possibly narrowed by cooling the sample e.g. in a supersonic beam or by collisions with a buffer gas.

Nonlinear mechanisms for canceling the Doppler effect have been harnessed with tunable lasers or with a single frequency comb. In direct frequency comb Doppler-free two-photon spectroscopy [1,2], all comb lines may contribute and the excitation probability of any given level can be the same as with a continuous-wave (cw) laser of the same average power. Short pulses facilitate nonlinear frequency conversion to access spectral regions where cw lasers are not readily available. However, the spectrum is only retrieved modulo the comb line spacing. Therefore, direct frequency comb spectroscopy with a single frequency comb is only suitable for spectra composed of very few transitions.

Here we propose and demonstrate Doppler-free Fourier transform spectroscopy by combining dual-comb spectroscopy and two-photon excitation in a standing-wave field. Dual-comb spectroscopy is a recent technique of Fourier spectroscopy without moving parts, where the time-domain interference between two combs of slightly different repetition frequencies is recorded. Mostly exploited for linear absorption spectroscopy, first proof-of-principle experiments have extended its potential to nonlinear phenomena [3]. Our new technique enhances such capabilities by combining sub-Doppler resolution with a broad spectral range.

We use two laser frequency combs (labeled 1 and 2) with slightly different repetition frequencies. For each laser, two trains of pulses counter-propagate in the sample to provide Doppler-free two-photon excitation. The excitation by pairs of counter-propagating pulse pairs produces Ramsey-like interference in the atomic excitation amplitude. The linearly increasing time separation between the pair of pulses of comb 1 and the pair of pulses of comb 2 automatically and continuously scans the interference phase. We measure such a time-domain interference by recording the modulations in the intensity of the fluorescence of the sample emitted during decays to lower states. Each excited transition is uniquely identified by the modulation imparted by the interfering excitations. The Fourier transform of the interferogram reveals a spectrum with a free-spectral range as broad as the span of the exciting laser combs. The resolution in one spectrum is determined by the comb line spacing. The position of the comb lines may be scanned in a step-wise manner and several spectra may be interleaved to improve the resolution possibly down to the optical comb line-width.

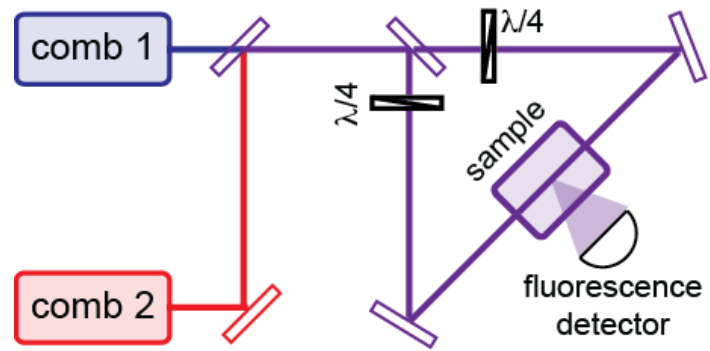

**Figure 1. Sketch of the experimental set-up**

In the experimental set-up (Fig.1), two amplified erbium-doped comb generators emit around 190 THz. Their repetition frequency is about 100 MHz and the difference in their repetition frequencies is 380 Hz. They are frequency-doubled using MgO:PPLN crystals to yield spectra centered around 385 THz that span 10 THz. The repetition frequencies are stabilized against the radio-frequency signal of a hydrogen maser, as all electronics in the set-up. One of their comb lines is stabilized to a 193 THz narrow-linewidth cw laser.

This cw laser is referenced to a metrological self-referenced frequency comb. The width of the optical comb lines around 195 THz is measured to be narrower than 1 kHz at 1 s. The beams of the two frequency-doubled combs are combined before entering an anti-resonant ring, which contains a heated rubidium cell. The two counter-propagating beams are circularly polarized with the same sense of rotation. They are focused at the same position of the cell and the total average power at the focus is about 10 mW, balanced between the four contributing comb beams (two in each direction). The fluorescence at 420 nm, induced by the two-photon excitation, is collected with a photomultiplier. The time-domain interference signal is recorded on a fast digitizer. The recording time per interferogram is 36 s, which corresponds to $67 \times 10^6$ samples. An a-posteriori correction scheme [4] corrects the interferograms for the residual relative fluctuations of the combs. To improve the resolution, limited in one spectrum by the comb line spacing of 100 MHz, we interleave one hundred spectra, each recorded by shifting the positions of the optical comb lines. As a result, we sample the Doppler-free line profiles of the two-photon transitions around 770 THz with an effective resolution of 1 MHz.

Figure 2 illustrates the resulting Doppler-free dual-comb spectrum of the 5*S*-5*D* transitions of $^{85}$Rb and $^{87}$Rb in natural abundance. The spectrum only plots the comb line positions of the interleaved spectrum. The main panel in Fig. 2 shows the entire optical domain where two-photon transitions are observed. Because rubidium has a relatively simple spectrum, the displayed span only covers 100 GHz, whereas the employed lasers simultaneously interrogate a region extending over 10 THz. Direct frequency comb spectroscopy with a single comb of 100-MHz line spacing would nevertheless fold all these transitions in a free spectral range of 100 MHz. At the pedestal of the intense $5S_{1/2}$-$5D_{5/2}$ lines around 770.57 THz, the residual Doppler profile can be distinguished. The signal-to-noise of the most intense transition ($5S_{1/2}$ *F*=3 - $5D_{5/2}$) of $^{85}$Rb approaches 12000. The good signal-to-noise ratio is a consequence of the efficient excitation scheme, where many pairs of comb lines contribute, and of the background-free fluorescence detection. The inset in Fig. 2 provides a magnified view of the $5S_{1/2}$ *F*=2 - $5D_{3/2}$ *F*=0,1,2,3 hyperfine transitions of $^{87}$Rb, surrounded in red in the main panel. The observed width of the lines is 6 MHz, due to the time-of-flight broadening. The frequency scale is directly referenced to the hydrogen maser. The experimental absolute transition frequencies of the non-blended lines are found to agree within 200 kHz with the literature. We attribute the main source of uncertainties to frequency pulling effects due to the absorption of the comb light by the intermediate 5*P* state before the excitation focal point in the cell. A detailed assessment of the systematic effects in our technique is beyond the scope of the present report and it will be performed in the near future with better-suited transitions in a more controlled sample environment. The experiments summarized above are, to our knowledge, the first report [5,6] of broadband partly-multiplex Doppler-free spectroscopy.

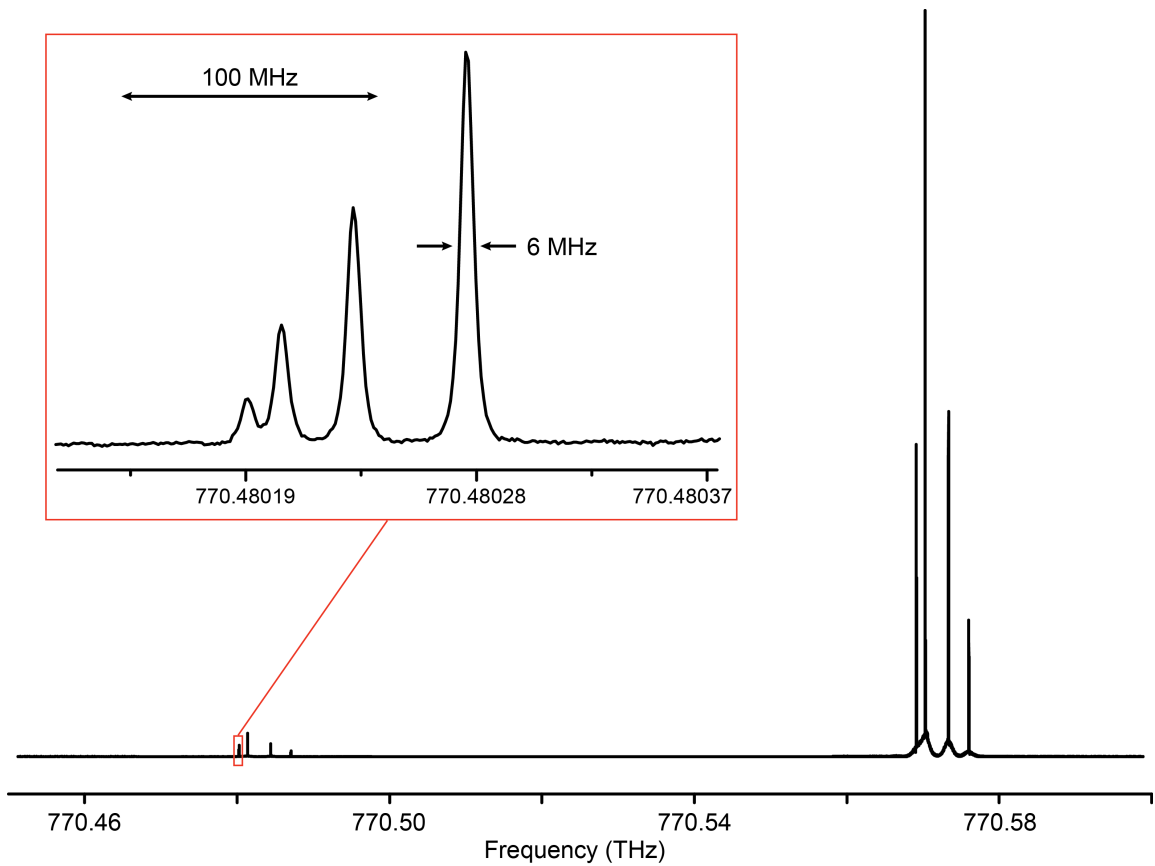

**Figure 2.  Experimental sub-Doppler dual-comb two-photon spectrum of the hyperfine transitions of $^{85}$Rb and $^{87}$Rb.**

For truly multiplex measurements, one would have to use lasers with a line spacing equal to the desired spectral resolution. Small comb line spacing, or low repetition frequencies, could take advantage of amplified laser systems. As an additional benefit, their high pulse energy would facilitate the nonlinear frequency conversion of the exciting light in the mid-infrared or ultraviolet (UV) spectral region and the nonlinear excitation of the sample. A significant challenge, especially in the UV range, may be to achieve narrow comb line width and high mutual coherence with amplified systems. Feed-forward stabilization might provide solutions.  The prospect of using amplified systems for dual-comb spectroscopy is consistent with the path followed with narrowband Ramsey comb spectroscopy [7] and opposite to that of direct frequency-comb two-photon spectroscopy with a single comb of a high repetition frequency, for which state-of-the-art systems still lack sufficient power [8]. Our new technique might open up novel approaches to precision two-photon spectroscopy in the UV or even XUV. More generally, the combination of a wide spectral bandwidth and sub-Doppler resolution may enable broadband molecular spectroscopy with unprecedented precision.

**Acknowledgments.** This research was financially supported by the European Research Council (Advanced Investigator Grant 267854), the Munich Center for Advanced Photonics and the Max-Planck Foundation.